\documentclass[aps,prl,twocolumn,superscriptaddress, showpacs]{revtex4}
\usepackage{graphicx}

\begin{document}

\title{Asymmetries in the electron-hole pair dynamics and strong Mott pseudogap effect in the phase diagram of cuprates}
\author{M.Z. Hasan}\altaffiliation{To whom correspondence should be addressed:
mzhasan@Princeton.edu} \affiliation{Department of Physics, Joseph
Henry Laboratories, Princeton University, Princeton, NJ
08544}\affiliation{CMC-CAT, Advanced Photon Source, Argonne
National Laboratory, Argonne, IL 08544}
\author{Y. Li}
\author{D. Qian}
\affiliation{Department of Physics, Joseph Henry Laboratories,
Princeton University, Princeton, NJ 08544}
\author{Y.-D. Chuang}
\affiliation{Advanced Light Source, Lawrence Berkeley National
Lab, Berkeley, Ca 94720}
\author{H. Eisaki}
\affiliation{AIST, 1-1-1 Central 2, Umezono, Tsukuba, Ibaraki,
305-8568 Japan}
\author{S. Uchida}
\affiliation{Department of Physics, University of Tokyo,  Tokyo
113-8656, Japan}
\author{Y. Kaga}
\author{T. Sasagawa}
\affiliation{Department of Adv. Materials Science, University of
Tokyo, Kashiwanoha, Chiba 277-8561, Japan  and CREST-JST,
Kawaguchi, Saitama 332-0012, Japan}
\author{H. Takagi}
\affiliation{AIST, 1-1-1 Central 2, Umezono, Tsukuba, Ibaraki,
305-8568 Japan}\affiliation{Department of Adv. Materials Science,
University of Tokyo, Kashiwanoha, Chiba 277-8561, Japan  and
CREST-JST, Kawaguchi, Saitama 332-0012, Japan}

\date{\today}


\begin{abstract}

Electron behavior in doped Mott systems continues to be a major
unsolved problem in quantum many-electron physics. A central issue
in understanding the universal Mott behavior in a variety of
complex systems is to understand the nature of their electron-hole
pair excitation modes. Using high resolution resonant inelastic
x-ray scattering (RIXS) we resolve the momentum dependence of
electron-hole pair excitations in two major classes of doped
copper oxides which reveals a Mott pseudogap over the full
Brillouin zone and over the entire phase diagram. The pair
bandwidth and zone-boundary pair velocity renormalize on either
sides of the phase diagram at very different rates indicating
strong particle-hole doping asymmetries observed for the first
time.
\end{abstract}


\maketitle

The discovery of high temperature superconductivity, colossal
magnetoresistance and other unusual electron transport behavior
has led to extensive research interests in doped Mott
insulators\cite{1}-\cite{7}. The unusual behavior of these systems
are often described by charge transport, optical excitation
spectrum or thermal and magnetic susceptibility measurements which
essentially probe the density of electron-hole pair excitation
modes, electronic charge or spin density of states. Transport or
thermodynamic measurements do not resolve the momentum quantum
numbers of pair excitations.
 Optical techniques which
are typically used to study the electronic excitation spectrum and
charge gap (Mott gap), are only confined near the zero momentum
transfer due to the large wavelength compared to the lattice
parameters\cite{8,9} . Resolving the pair excitations along the
momentum axis provides a new and important dimension of
information to understand the electron dynamics in complex
systems.

A parent Mott insulator can be doped either with electrons or
holes, and many physical properties such as transport and
magnetism are very sensitive to the doping level. Understanding
the doping evolution of the electronic structure in both
dopant-sides is quite important. Electronic structure of cuprates
has been extensively studied by Angle-resolved Photoemssion
spectroscopy (ARPES). However, ARPES provides key insights into
the single particle dynamics only. Using high resolution resonant
inelastic x-ray scattering \cite{10,11}we resolve the momentum
dependence of electron-hole pair excitations in two major classes
of copper oxide series including Nd$_{2-x}$Ce$_{x}$CuO$_4$ and
La$_{2-x}$Sr$_{x}$CuO$_4$, systematically over the phase diagram
to gain insights into the universal aspects of correlated electron
motion in these systems.

The inelastic x-ray scattering data were obtained at the CMC-CAT
beamline 9-ID and BESSRC-CAT beamline 12-ID at the Advanced Photon
Source, Argonne National Laboratory.  The incident beam energy was
selected near the copper K edge to produce enhancement of
inelastic signals. The scattered beam was reflected off of a
germanium (733) crystal analyzer and measured with a solid-state
detector. The overall energy resolution for the experiment was 370
meV and all the data were taken at room temperature. Experimental
details of scattering parameters are described in ref. [12,13].
Incident x-ray polarization was along the \it c\rm-axis of all
samples
 in a vertical scattering geometry. Furthermore, scattering was also performed
in the horizontal geometry where the polarization was within the
\it ab\rm-plane to check for matrix element effects possibly
suppressing intensity near the leading-edge of the gap. Data were
taken with closely spaced momentum intervals in between
\textit{\textbf{q}} \rm$\sim$  $\pi$ to $2\pi$ (
\textit{\textbf{q}} being the scattering wave vector) to allow for
leading-edge mid-point analysis of the gap for the first time.

Fig.1(top) shows pseudo-color plots of momentum-resolved charge
excitation spectra in the electron-doped cuprate series
Nd$_{2-x}$Ce$_x$CuO$_4$ (e-doped, x=0.10, 0.13) (a), hole doped
cuprtae series La$_{2-x}$Sr$_x$CuO$_4$ (h-doped, x=0.02, 0.14) (b)
and undoped cuprate (x=0) (c) as a function of hole and electron
doping (x) level. Quasielastic scattering was removed by fitting
below 0.8 eV. A broad excitation band (cyan) is seen from 0.5 to 3
eV for the e-doped systems whereas the analogous excitations in
the h-doped systems (red) appear in 2 to 5 eV energy range. These
excitations are seen over the range of momentum transfers covering
the full Brillouin zone. Fig.1(d) is typical plot for
momentum-resolved energy-loss spectra corresponding to the image
in Fig.1(a)/left.
\begin{figure}[ht] \center
\includegraphics[width=8.0cm]{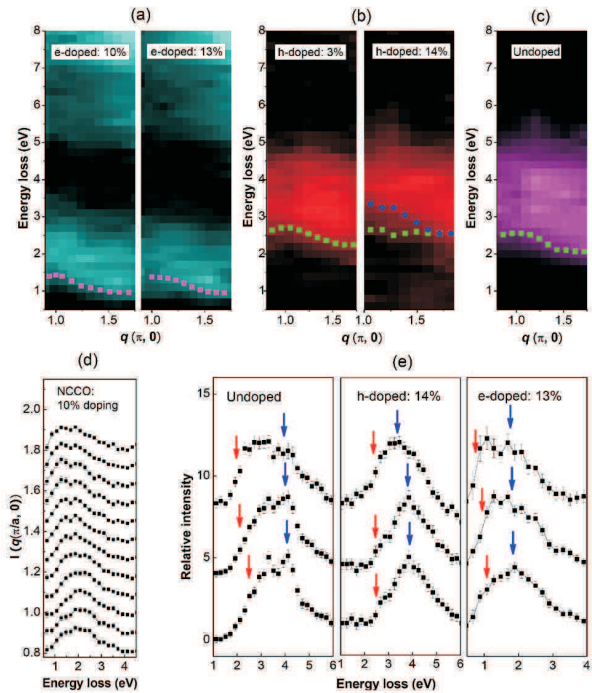}\caption{Momentum-Resolved Charge
Excitations in Cuprates: Momentum
($\hbar$\textit{\textbf{q}})-dependence of charge excitations in
electron-doped Nd$_{2-x}$Ce$_x$CuO$_4$ (a), hole-doped
La$_{2-x}$Sr$_x$CuO$_4$ (b) and undoped insulator (c) copper oxides. Brighter color
indicates higher intensity. Dotted lines indicate the location of
the leading edge of the gap except in case of LSCO(x=0.14) where a
second dotted line traces out the second peak. (d) plots the \textit{\textbf{q}}-resolved energy-loss
spectra corresponding to the image plot (a)/left. (e) shows selected energy-loss
curves near high symmetry points of the Brillouin zone for undoped, \it h\rm -doped, and \it e\rm -doped
samples }
\end{figure}
Fig.1(e) shows selected energy-loss spectra near high symmetry
points of the Brillouin zone to compare the effect of hole and
electron doping by comparing insulator vs. metal. The lower energy
edge (dotted lines in Fig.1(top)) of the dominant excitation
defines the edge of the insulating charge gap (effective Mott gap)
[in the insulator (Fig.1c)]. For the undoped sample (Fig.1c), near
the zone center the onset of excitation edge is at the lowest in
energy at about 1.8 eV, consistent with optical studies \cite{8}
where as the analogous excitation in NCCO series appears at around
1 eV, again consistent with optical studies \cite{9}. The
\textit{\textbf{leading-edge}} mid-point of the gap observed here
hardly disperses in going from the zone center
(\textit{\textbf{q}} $\sim$ 2$\pi$) to \textit{\textbf{q}} $\sim
1.4\pi$ (unlike earlier reports on La$_2$CuO$_4$ \cite{14,15} and
a concurrent work on LSCO \cite{16}). However, this behavior we
observe is consistent with dispersions seen in other cuprates such
as Ca$_2$CuO$_2$Cl$_2$\cite{12} and Sr$_2$CuO$_2$Cl$_2$\cite{14}
and the Nd$_{2-x}$Ce$_x$CuO$_4$ series studied here.

With increased doping the spectra change in two major ways - a
low-energy continuum appears below the gap as seen in Fig.2 and
the momentum dependence of the leading edge weakens. A closer look
in Fig.2 shows that the low-energy spectral weight grows
(reddish/bluish continuum in Fig.1a and Fig.1b) at the expense of
the spectral weight around 2 to 3.5 eV range in the insulator for
hole doping and around 1 to 2 eV range for electron doping
(Fig.2). This is clear evidence that even if doping introduces
states in the middle of the gap (as argued based on ARPES
measurements\cite{17,18}), changes in the electronic structure are
further accompanied by direct melting (clear systematic changes in
the leading-edge behavior) of the Mott gap in a momentum dependent
manner. The observed low-energy continuum in the doped system is
due to pair excitations involving quasiparticle-quasihole states
that appear in the middle of the gap upon doping. In ARPES, only
single electron state components of this continuum are seen as
low-energy quasiparticle states\cite{14}. The spectral
distribution of the continuum is momentum dependent - changes upon
doping are weak near the zone boundary (\textit{\textbf{q}} $\sim$
($\pi$,0)) as opposed to the zone center (\textit{\textbf{q}}
$\sim$ (2$\pi$,0)) (Fig.2). Fermi surface evolves asymmetrically
by electron vs. hole doping and the differences in the detail
momentum dependence of Mott gap melting in electron vs. hole
doping would then be related to different topologies of Fermi
surfaces in LSCO and NCCO as observed in ARPES. Quantitative
analysis of the lower energy continuum weight below 0.8 eV is
difficult due to the large quasielastic tails from structural
diffuse scattering.

We focus rather on the spectral-edge behavior of the charge gap
since low-energy weight is generated at the expense of the weight
near the gap (Fig.2) and provides a connection or correlation
(related to the spectral weight transfer) to the low-energy
physics of these systems. A closer look at the gap in doped
systems is interesting in its own right since a hierarchy of
increasingly higher energy scales seems to be an emergent theme in
the physics of doped Mott insulators\cite{4}-\cite{6}. Our results
show that a gap feature similar to the undoped insulator exists at
all doping and its dispersion is a remnant behavior from the
insulator at all momenta as seen from Fig.1 and 2. We refer to
this feature in the doped system as the Mott pseudogap or remnant
Mott gap. A remnant Fermi surface-like behavior of single-particle
spectral weight, \textit{n(k)}, is observed in the undoped Mott
insulator in ARPES studies by integrating within an energy window
to include the occupied band \cite{19}. In inelastic x-ray
spectrum both occupied and unoccupied bands contribute hence our
observation of remnant Mott gap in the doped system with similar
momentum dispersion as in the insulator is consistent with a
remnant Fermi surface (momentum behavior) behavior.
\begin{figure}[t]
\center \includegraphics[width=5.3cm]{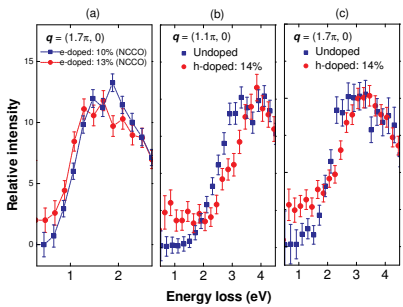} \caption{Doping
dependence of low-energy spectral weight with momenta near zone
boundary for electron doping (left) and near zone center (middle)
and near zone boundary (right) for hole doping.}
\end{figure}

In an energy-loss experiment one measures particle-hole pair modes
in metals and insulators and damped plasmon modes in metals. The
dynamics and nature of particle-hole pair ("excitons") in Mott
insulators is less understood. The difficulty in describing the
pair dynamics in a Mott insulator arises because of the fact that
where as in a band insulator, excitons form via Coulomb
interaction and the gap is due to Pauli exclusion, in Mott
insulators the gap itself has a direct Coulombic
origin\cite{20}-\cite{26}. However, this provides a possible
connection between the particle-hole pair spectrum in doped Mott
insulators to the charge-gap of the undoped insulator. When these
insulators are doped eventually a weak Drude response
emerges\cite{8,9} with corresponding plasma resonances being
typically well below 1 eV making it difficult to observe in
resonant x-ray experiments.

Fig.3(a) shows the energy vs. momentum behavior of the excitations
edges for both \textit{e}-doped and \textit{h}-doped systems. We
determine the leading-edge mid-point by taking a derivative of the
energy-loss spectra with respect to the energy-loss axis. Undoped
insulator shows the largest amount of dispersion which is about
500 meV. Upon hole doping dispersion seems to flatten out in a
systematic manner. For the 3\% \textit{h}-doping (insulator)
dispersion is reduced and follows a qualitative trend as seen in
the undoped insulator. In the 14\% \textit{h}-doped sample
dispersion is much reduced - it has flattened out quite a bit and
is on the order of 100-150 meV. We also note that for all dopings,
excitation-edge seem to be roughly pinned at the zone boundary
\textit{\textbf{q}} $\sim$ ($\pi$, 0) (Fig.3a).

\begin{figure}[t]
\includegraphics[width=8.6cm]{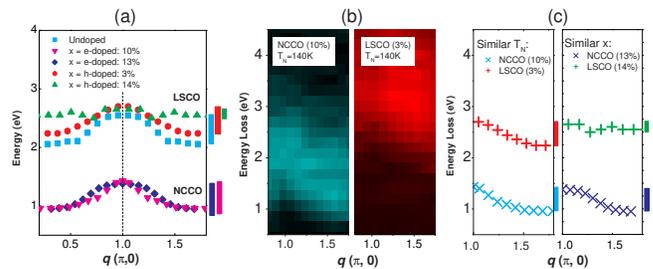} \caption{Pair Bandwidth and Magnetism :
(a) Energy vs. momentum relation of the leading-edge of
the Mott gap determined by taking first derivatives of the image
plots in Fig.1. Vertical bars outside the panel give a measure of
the bandwidth.(b) Comparison of excitation spectra with different doping but same
Neel temperature. (c) Comparison of pair dispersion for samples with
same Neel temperature (left) and for samples with same doping but
different Neel temperature (right) on either side of the phase
diagram.}
\end{figure}

\begin{figure*}[ht]
\includegraphics[width=12cm]{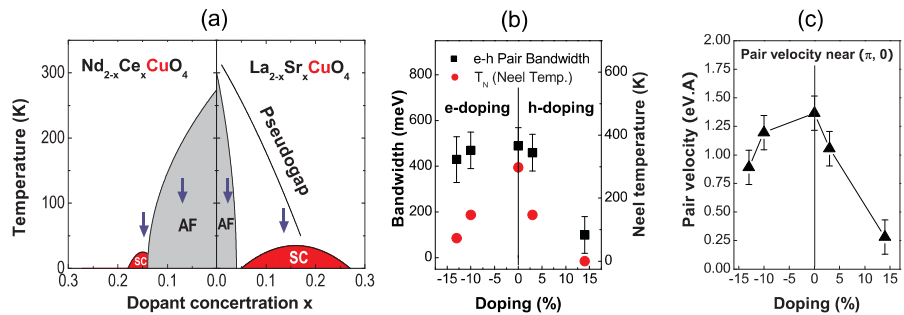} \caption{Neel Order, Pair Bandwidth and Zone-Edge Velocity:
(a). Electronic phase diagram of doped copper oxides. (b). Doping
dependence of electron-hole pair bandwidth and Neel temperature.
(c). Doping dependence of zone boundary ($\pi$, 0) pair velocity.}
\end{figure*}

Based on simple electronic structure considerations, doping of a
few holes into two dimensional insulating cuprates creates a
Zhang-Rice singlet state whose wavefunction is spread over four
neighboring oxygen atoms whereas the doping of electron creates
states in the upper Hubbard band whose character (wavefunction) is
much more \textit{d}-like with strongly localized
character\cite{20}-\cite{23}. However, the way these states evolve
to the high doping case is not well understood. Numerical
finite-lattice simulations suggest asymmetric evolution of
electron-hole pair modes in doped Mott insulators\cite{22}. Our
results suggest a broad agreement with such simulations. However,
a detailed comparison with lattice simulations are difficult since
finite size of the lattice allows for only a few points to be
studied. Much dramatic changes due to hole doping may be
understood by combining ARPES results and Hubbard model spectral
functions for electron-hole pair excitations. As evident from
Fig.3(a), dispersion of pair excitations is weak from the zone
center to about the half-way (1.5$\pi$, 0) to the zone boundary.
In comparison with lattice simulations this effect has been argued
to be due to the fact that the lowest energy state of the upper
Hubbard band is near ($\pi$, 0) as opposed to the momentum of
($\pi$/2, $\pi$/2) of the top-most state of the lower Hubbard
band\cite{12,22,23}. Under this scenario, the x-ray spectra in
lightly doped NCCO should not be much different from the undoped
insulator since in the electron doped system lowest-energy excited
electrons are in the upper Hubbard band and have very similar
momenta as in the undoped insulator whereas in LSCO the lowest
energy excited electrons are still in the lower (Zhang-Rice)
Hubbard band. In other words, in case of hole doping, upper
Hubbard band is not involved in creating the low energy
electron-hole pair excitations hence the pair dynamics involves
states of very different characters.

Fig.3(b) and (c) show a comparison of charge excitation response
in 3\%-LSCO and 10\%-NCCO. Doping levels are picked so that both
compounds have about the same Neel temperature ($\sim$ 140K).
Fig.3(b) shows the dispersion behavior of the leading-edge. Apart
from an off-set in the energy scale (electron doped system has a
smaller optical gap $\sim$ 1.1 eV\cite{9}) both compounds exhibit
qualitatively similar momentum behavior and have the same amount
of energy dispersion or pair bandwidth ($\sim$ 300 meV). This is
on the order of 2\textit{\textbf{J}} to 3\textit{\textbf{J}}
(where \textit{\textbf{J}} is the exchange constant measured by
neutron scattering (27)) which is a relevant magnetic energy scale
for these systems. Fig.3(c) shows a comparison of LSCO(14\%) and
NCCO(13\%) which have very similar doping levels. In the case of
electron-doping (NCCO), the dispersive edge is much robust whereas
a similar amount of hole doping leads to much \bf more \rm
renormalization of the dispersion. Even for the 13\% electron
doped metallic NCCO dispersion remains on the order of 300 meV.
This is a factor of two larger than the dispersion seen in the
optimal hole doping. This is an indication of the dominance of
antiferromagnetic correlations in NCCO which is consistent with
findings in neutron studies of NCCO\cite{28}.

We look for possible direct correlation of electron-hole pair
bandwidth measured in x-ray scattering and the Neel temperature
which is the energy scale for the onset of long-range
antiferromagnetic order in these systems. The cuprate phase
diagram is shown in Fig.4(a). Fig.4(b) plots the pair bandwidth
and Neel temperature as a function of doping. This correlation
suggests that pair bandwidth decreases as long-range magnetic
correlations decrease with doping. However, the collapse of
bandwidth is much faster with hole doping than it is for electron
doping. For the hole doping, pair bandwidth shows a direct scaling
with Neel temperature which suggests a direct correlation with
antiferromagnetic order in the hole doped systems. In case of
electron doping bandwidth suppression with doping is weaker and
weakly correlated with the doping fall-off of antiferromagnetic
order. Robustness of strong short-range antiferromagnetic order is
clearly seen in neutron scattering studies of highly doped NCCO
systems\cite{28}. We further introduce a velocity : the derivative
of the excitation band near ($\pi$, 0) which can be thought of as
the zone-boundary pair velocity ($\delta\omega/\delta q$). Such
velocity exhibits similar correlation with doping. Like bandwidth,
pair velocity is also highly asymmetric in electron vs. hole
doping (Fig.4c). These results taken together, suggest that
coherent propagation of electron-hole pairs are largely affected
by the effective magnetic coupling or interaction strengths and
reflects a strong electron-hole doping asymmetry in the electron
dynamics of cuprates.

Our results reveal momentum structure of remnant Mott gap and
electron-hole pair excitations in doped cuprates and their
correlation with the phase diagram for the first time (Fig.4).
Although much theoretical efforts are needed to understand these
observations we report here, results by themselves shed new light
on the theories that argue for an intimate connection between the
pseudogap and Mott charge gap and the importance of magnetic
coupling strength in describing correlated electron behavior in
cuprates\cite{3,29}.

We   acknowledge  M. Beno, D. Casa and T. Gog for technical help.
The experiments were performed at the Adv. Photon Source of
Argonne National Lab which is supported by the U.S. Dept. of
Energy, BES under Contract No. W-31-109-ENG-38. This work was
partially supported by an NSF (DMR-0213706) grant. MZH
acknowledges partial support through R.H. Dicke research award.

\end{document}